\documentclass[aps,prl,twocolumn,epsfig,groupedaddress]{revtex4}

\usepackage{epsfig}
\usepackage{amsmath}
\usepackage{graphicx}
\usepackage{bm}

\graphicspath{{Figs/}}

\newcommand{\be}{\begin{equation}}
\newcommand{\ee}{\end{equation}}
\newcommand{\bea}{\begin{eqnarray}}
\newcommand{\eea}{\end{eqnarray}}
\newcommand{\bem}{\begin{multline}}
\newcommand{\eem}{\end{multline}}
\newcommand{\beg}{\begin{gather}}
\newcommand{\eeg}{\end{gather}}

\def\eq#1{{Eq.~(\ref{#1})}}
\def\fig#1{{Fig.~\ref{#1}}}
\newcommand{\ben}{\begin{eqnarray*}}
\newcommand{\een}{\end{eqnarray*}}

\begin{document}

\title{Heavy Quark Potential at Finite Temperature Using the
  Holographic Correspondence} 
\author{Javier L. Albacete} \affiliation{Department of
  Physics, The Ohio State University,
  Columbus, OH 43210, USA}
\author{Yuri V. Kovchegov}
\email[Correspond to\ ]{yuri@mps.ohio-state.edu}
\affiliation{Department of Physics, The Ohio State University, 
  Columbus, OH 43210, USA}
\author{Anastasios Taliotis}
\affiliation{Department of Physics, The Ohio State University, 
  Columbus, OH 43210, USA}

\begin{abstract}
  We revisit the calculation of a heavy quark potential in ${\cal N}
  =4$ supersymmetric Yang-Mills theory at finite temperature using the
  AdS/CFT correspondence. As is widely known, the potential calculated
  in the pioneering works of Rey et al. \cite{Rey:1998bq} and
  Brandhuber et al. \cite{Brandhuber:1998bs} is zero for separation
  distances $r$ between the quark and the anti-quark above a certain
  critical separation, at which the potential has a kink.  We point
  out that by analytically continuing the string configurations into
  the complex plane, and using a slightly different renormalization
  subtraction, one obtains a smooth non-zero (negative definite)
  potential without a kink.  The obtained potential also has a
  non-zero imaginary (absorptive) part for separations $r > r_c =
  0.870/\pi T$ . Most importantly at large separations $r$ the real part of the
  potential does not exhibit the exponential Debye falloff expected
  from perturbation theory and instead falls off as a power law,
  proportional to $1/r^4$ for $r > r_0 = 2.702 / \pi T$.
\end{abstract}

\date{\today}


\maketitle


Heavy quark potential is a very important quantity in gauge theories
at finite temperature. It also has a great phenomenological relevance
in connection with experimental programs in heavy ion collisions at
the Relativistic Heavy Ion Collider and at the upcoming Large Hadron
Collider. The melting of heavy mesons in a medium is considered to be
one of the main experimental signatures for Quark-Gluon Plasma
formation, the ultimate goal of such experiments. Current analyses of
available experimental data indicate that the matter formed in such
collisions is strongly coupled. Thus, the study of heavy quark
potential requires strong-coupling techniques, such as the Anti-de
Sitter space/conformal field theory (AdS/CFT) correspondence
\cite{Maldacena:1997re,Gubser:1998bc,Witten:1998qj,Aharony:1999ti}.
The main goal of this work is to improve the current description of
heavy quark potential at finite temperature in the AdS/CFT framework.

Until recently heavy quark potential has been calculated either
analytically at small coupling using the perturbation theory, or
numerically using lattice simulations. With the advent of AdS/CFT
correspondence
\cite{Maldacena:1997re,Gubser:1998bc,Witten:1998qj,Aharony:1999ti}, it
became possible to analytically calculate the heavy quark potential at
strong coupling, albeit only for the ${\cal N} =4$ supersymmetric
Yang-Mills (SYM) theory.

The first calculation of a heavy quark potential in vacuum for ${\cal
  N} =4$ SYM theory was carried out by Maldacena in
\cite{Maldacena:1998im}. Soon after \cite{Maldacena:1998im}
calculations of the heavy quark potential for ${\cal N} =4$ SYM theory
at finite temperature appeared in \cite{Rey:1998bq,Brandhuber:1998bs}.
In \cite{Maldacena:1998im} the heavy quark potential is obtained from
the expectation value of a static temporal Wilson loop and in
\cite{Rey:1998bq,Brandhuber:1998bs} from the correlator of two
Polyakov loops. They are calculated by extremizing the world-sheet of
an open string attached to the quark and anti-quark located at the
edge of the AdS$_5$ space in the background of the empty AdS$_5$ space
in \cite{Maldacena:1998im} and in the background of the AdS$_5$ black
hole metric in \cite{Rey:1998bq,Brandhuber:1998bs}.

The zero-temperature heavy quark potential obtained in
\cite{Maldacena:1998im} is of Coulomb type due to conformal invariance
of ${\cal N} =4$ SYM theory:
\begin{align}\label{zeroT}
  V_0 (r) \, = \, - \frac{\sqrt{\lambda}}{2 \, \pi \, c_0^2 \, r}
\end{align}
with $\lambda$ the 't Hooft coupling and $c_0 \, = \, \Gamma^2
\left(\frac{1}{4}\right)/(2 \, \pi)^{3/2}$. Here $r$ is the distance
between the quark and the anti-quark in the boundary gauge theory.

The finite-temperature heavy quark potential obtained in
\cite{Rey:1998bq,Brandhuber:1998bs} starts out at small $r$ being
close to the vacuum potential of \eq{zeroT}, but rises steeper than
the vacuum potential, becoming zero at a separation $r^* = 0.754 /\pi
T$. For larger separations, i.e., for $r>r^*$, the authors of
\cite{Rey:1998bq,Brandhuber:1998bs} argue that the string ``melts'',
and the dominant configuration corresponds to two straight strings
stretching from the quark and the anti-quark down to the black hole
horizon. The resulting potential is thus zero for $r>r^*$ and has a
kink (a discontinuity in its derivative) at $r = r^*$.

To clarify the definition of the heavy quark potential at finite
temperature T let us first define the Polyakov loop for $SU(N_c)$
gauge theory at spatial location $\vec r$ by
\begin{align}\label{L}
  L ({\vec r}) \, = \, \frac{1}{N_c} \, \text{Tr} \left[ \text{P} \exp
    \left( i g \int_0^\beta d \tau \, A_4 ({\vec r}, \tau )\right)
  \right]
\end{align}
with $\tau$ the Euclidean time and $\beta = 1/T$. The connected
correlator of two Polyakov loops can be written as
\cite{McLerran:1981pb,Nadkarni:1986cz}
\begin{align}\label{LL}
  \langle L (0) \, L^\dagger ({\vec r}) \rangle_c = \frac{e^{- \beta
      \, V_1 (r) } + (N_c^2 -1) \, e^{- \beta \, V_{adj} (r)
    }}{N_c^2}.
\end{align}
\eq{LL} is the definition of singlet $V_1 (r)$ and adjoint $V_{adj}
(r)$ potentials in Euclidean time formalism. With the appropriate
modification of \eq{L}, \eq{LL} also applies to ${\cal N}=4$ SYM.

\begin{figure}[h]
\centering
\includegraphics[width=6.3cm]{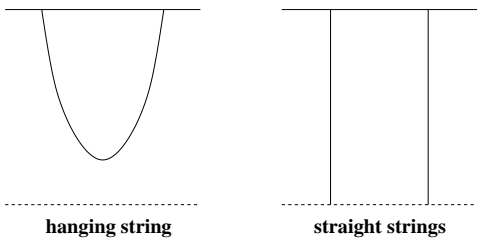}
\caption{Two configurations of open strings corresponding to 
  Polyakov loop correlator. Solid horizontal line denotes the boundary
  of the AdS space, while the dashed line denotes the location of the
  black hole horizon.~\vspace*{-3mm}}
\label{strings}
\end{figure}

To calculate the Polyakov loop correlator in AdS space one follows the
standard prescription outlined in
\cite{Maldacena:1998im,Rey:1998bq,Brandhuber:1998bs} and connects open
string(s) to the positions of Polyakov loops at the boundary of the
AdS space in all possible ways. The two relevant configurations are
shown in \fig{strings} and labeled ``hanging string'' and ``straight
strings''.

The two string configurations shown in \fig{strings} give two
different saddle points of the Nambu-Goto action. In the large-$N_c$
large-$\lambda$ limit the integral over all string configurations, and
hence the Polyakov loop correlator as well, is equal to the sum of the
contributions of the different saddle points. Actually, as was argued
in \cite{Bak:2007fk}, the two straight strings on the right of
\fig{strings} have Chan-Paton labels indicating which D3 brane each
string ends on. Therefore the straight strings configuration actually
represents of the order of $N_c^2$ extrema corresponding to the
different ways the two straight strings connect to $N_c$ D3
branes. Summing over all the
saddle points we write (in Euclidean space)
\begin{align}\label{LL2}
  \langle L (0) \, L^\dagger ({\vec r}) \rangle_c \propto \frac{e^{-
    S_{NG}^{hanging}} + (N_c^2-1) \, e^{- S_{NG}^{straight}}}{N_c^2}
\end{align}
with $S_{NG}^{hanging}$ and $S_{NG}^{straight}$ the Nambu-Goto actions
of the hanging and straight string configurations.

Comparing \eq{LL2} with \eq{LL} we conclude that the hanging string
configuration gives $V_1 (r)$, while the two straight strings
stretching to the horizon give $V_{adj} (r)$.  
However $V_{adj}$ itself is $N_c^2$-suppressed and repulsive, while
$V_1$ is of order one in $N_c$-counting and attractive
\cite{Nadkarni:1986cz}. Renormalizing the Nambu-Goto actions in
\eq{LL2} by subtracting the actions of the string configurations at
infinite quark--antiquark separations one obtains $S_{NG,
  ren}^{straight} =0$, which implies that $V_{adj}$ is zero at leading
order in $N_c^2$.  The first non-trivial contribution to $V_{adj}$ is
given by graviton exchanges between the strings in the bulk calculated
in \cite{Bak:2007fk}. If exponentiated (eikonalized), they would
indeed give $N_c^2$-suppressed contributions in the exponent, as
expected for $V_{adj} (r)$. Here we will calculate the singlet
potential $V_1 (r)$. In lattice simulations it is usually $V_1 (r)$
which is understood as the heavy quark potential at finite temperature
\cite{Kapusta}. In the real-time formalism $V_1 (r)$ is given by the
expectation value of a static (temporal) Wilson loop via
\begin{align}\label{Wilson}
  \langle W \rangle \, = \, e^{- i \, {\cal T} \, V_1 (r)}
\end{align}
with the temporal extent of the Wilson loop $\cal T \rightarrow
\infty$. Note that when calculating the Wilson loop (\ref{Wilson}) in
Minkowski space only the hanging string configuration contributes, as
the quark and the anti-quark are projected onto a color-singlet state
at initial and final times.

To find $V_1 (r)$ (henceforth referred to as $V (r)$) we will study
the behavior of the hanging string solution found in
\cite{Rey:1998bq,Brandhuber:1998bs} for $r>r^*$. As is well-known, for
$r > r_c = 0.870/\pi T$ the string coordinates of the solution
\cite{Rey:1998bq,Brandhuber:1998bs} become complex-valued. This
  simply indicates that the saddle point of the Nambu-Goto action lies
  in the complex string coordinate region: it does not invalidate the
  saddle point approximation and the results obtained with it. 
Similar complex-valued solutions were recently observed by the authors
in \cite{Albacete:2008ze}, where the scattering amplitude of a
quark--anti-quark dipole on a shock wave was calculated. In
\cite{Albacete:2008ze} the complex-valued string coordinates were
instrumental for finding the unitary solution for the scattering cross
section. Inspired by that example, below we will analytically continue
the potential of \cite{Rey:1998bq,Brandhuber:1998bs} into the complex
region of string coordinates. The resulting potential is smooth. The
corresponding force on the quarks is a continuous function of $r$. By
modifying the ultraviolet (UV) subtraction we obtain a potential which
is non-zero for all separations $r$. The potential develops an
imaginary part, corresponding to the decay of the quark--anti-quark
singlet state: similar results have been seen in finite temperature
perturbation theory in \cite{Laine:2006ns,Brambilla:2008cx}. Finally,
instead of Debye screening leading to exponential falloff of the
potential at large distances, we find the power-law falloff $\text{Re}
[V(r)] \sim 1/r^4$ at large $r$.

We want to calculate a temporal Wilson loop in a finite-temperature
${\cal N} =4$ SYM medium. We shall define the real-time heavy quark
potential in the same way as in \cite{Laine:2006ns}.  Following
\cite{Rey:1998bq,Brandhuber:1998bs} we start with the AdS$_5$ black
hole metric in Minkowski space
\cite{Witten:1998zw,Maldacena:1997re,Janik:2005zt}
\begin{align}
  ds^2 = \frac{L^2}{z^2} \left[ - \left( 1 - \frac{z^4}{z_h^4} \right)
    dt^2 + d {\vec x}^2 + \frac{d z^2}{1 - \frac{z^4}{z_h^4}} \right]
\end{align}
where $d {\vec x}^2 = (d x^1)^2 + (d x^2)^2 +(d x^3)^2$, $z$ is the
coordinate describing the 5th dimension and $L$ is the curvature of
the AdS$_5$ space. The horizon of the black hole is located at $z =
z_h$ with $z_h = 1/\pi T$.

We want to extremize the open string worldsheet for a string attached
to a static quark at $x^1 = r/2, x^2 = x^3 =0$ and an anti-quark at
$x^1 = -r/2, x^2 = x^3 =0$. Parameterizing the static string
coordinates by
\begin{align}
  X^\mu \! = \! \left[ X^0=t, X^1=x , X^2=0, X^3=0, X^4=z(x) \right]
\end{align}
we write the Nambu-Goto action as
\begin{align}\label{NG1}
  S_{NG} (r, T) \, = \, - \frac{\sqrt{\lambda}}{2 \, \pi} \, {\cal T}
  \int_{-r/2}^{r/2} dx \, \sqrt{\frac{1+z'^2}{z^4} - \frac{1}{z_h^4}},
\end{align}
where $z' = d z(x) /dx$.

The Euler-Lagrange equation corresponding to the action (\ref{NG1}) is
\begin{align}\label{EL}
  (2 + z \, z'') \, (z^4 - z_h^4) - 2 \, z'^2 \, (z^4 + z_h^4) =0
\end{align}
with $z'' = d^2 z/dx^2$. Solving \eq{EL} with the boundary conditions
$z (x=\pm r/2) =0$ one gets
\begin{align}\label{sol}
  x + \frac{r}{2} \, = \, \frac{z^3}{3 \, z_h^2 \, z_{max}^2} \, &
  \sqrt{z_h^4 - z_{max}^4} \notag \\ \times \, & F_1 \left(
    \frac{3}{4} ; \frac{1}{2}, \frac{1}{2}; \frac{7}{4} ;
    \frac{z^4}{z_h^4}, \frac{z^4}{z_{max}^4}\right)
\end{align}
where $F_1$ is the Appell hypergeometric function. Here $z_{max}$ is
the constant of integration corresponding to the maximum of the string
coordinate along the fifth dimension of the AdS$_5$ space, whose
boundary is located at $z=0$. It is given by the solution of the
following equation
\begin{align}\label{zmax}
  r \, c_0 \, = \, \frac{z_{max}}{z_h^2} \, \sqrt{z_h^4 - z_{max}^4} \ 
  \ F \left( \frac{1}{2}, \frac{3}{4} ; \frac{5}{4} ;
    \frac{z_{max}^4}{z_h^4} \right)
\end{align}
with $F$ the hypergeometric function.  Indeed in the $T \rightarrow 0$
limit $z_h \rightarrow \infty$ and \eq{zmax} gives us $z_{max} = r
c_0$ in agreement with Maldacena's vacuum solution
\cite{Maldacena:1998im}.

The action in \eq{NG1} contains a UV divergence, which has to be
subtracted out. Usually, the subtraction contains a finite piece as
well \cite{Maldacena:1998im,Rey:1998bq,Brandhuber:1998bs}, which may
be temperature-dependent in the case at hand. Here we will use the
following subtraction, different from the one used in
\cite{Rey:1998bq,Brandhuber:1998bs}: we define the quark--anti-quark
potential by
\begin{align}\label{subtr}
  V (r) \, = \, - \left\{ S_{NG} (r, T) - \text{Re} [S_{NG} (r =
    \infty, T)] \right\} / {\cal T}.
\end{align}
This subtraction insures that the real part of the potential $V(r)$
goes to zero at infinite separations. Our subtraction (\ref{subtr}) is
consistent with that used in \cite{Maldacena:1998im} to find the heavy
quark potential at zero temperature.

Using the solution from \eq{sol} in Eqs. (\ref{NG1}) and (\ref{subtr})
we obtain the following expression for the heavy quark potential of
the ${\cal N} =4$ SYM theory at finite temperature:
\begin{align}\label{pot}
  V (r) \, = \, \frac{\sqrt{\lambda}}{2 \, c_0 \, \pi} \, \bigg[ - &
  \frac{1}{z_{max}} \, \left( 1 - \frac{z_{max}^4}{z_h^4} \right)
  \notag \\ \times \, & F \left( \frac{1}{2}, \frac{3}{4} ;
    \frac{1}{4} ; \frac{z_{max}^4}{z_h^4} \right) + \frac{1}{z_h}
  \bigg].
\end{align} 
\eq{zexp} below is also needed to obtain \eq{pot}.  Our subtraction
prescription resulted in the $1/z_h$ term on the right of \eq{pot}
instead of $2 c_0 /z_h$, which would correspond to the subtraction
done in \cite{Rey:1998bq,Brandhuber:1998bs}. \eq{pot} along with
\eq{zmax} gives us the heavy quark potential as a function of the
separation $r$ and temperature $T = 1/\pi z_h$.

As can be readily checked numerically, $z_{max}$ given by \eq{zmax}
becomes complex for $r > r_c = 0.870 \, z_h$, leading to
complex-valued $z(x)$ and the potential $V(r)$. This led the authors
of \cite{Rey:1998bq,Brandhuber:1998bs} to abandon their solution for
$r > r_c$ (in fact, the solution was abandoned even earlier, for $r >
r^*$). We suggest however to interpret the complex-valued saddle
points as corresponding to quasi-classical configurations in the
classically forbidden region of string coordinates.  This is similar
to the method of complex trajectories used in quasi-classical
approximations to quantum mechanics \cite{LL3}.  

The complexification of the string coordinates simply indicates that
the saddle point of the integral over string coordinates becomes
complex. According to the standard AdS/CFT prescription
\cite{Maldacena:1998im}, in the large-$N_c$ large-$\lambda$ limit the
integral over string coordinates is still dominated by the saddle
point, even if it is complex. Therefore, the fact that string
coordinates at the saddle point become complex does not undermine the
validity of the approximation.  Below we will extend the solution of
Eqs.  (\ref{pot}) and (\ref{zmax}) to $r > r_c$ allowing for
complex-valued $z_{max}$.

\begin{figure}[h]
\centering
\includegraphics[width=6.3cm]{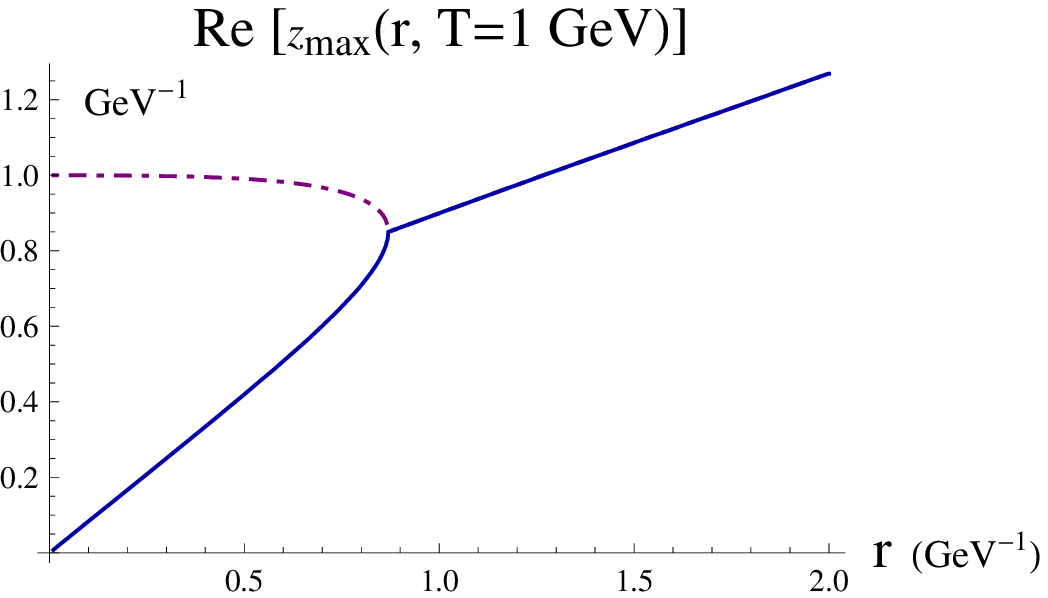} \\~\\
\includegraphics[width=6.3cm]{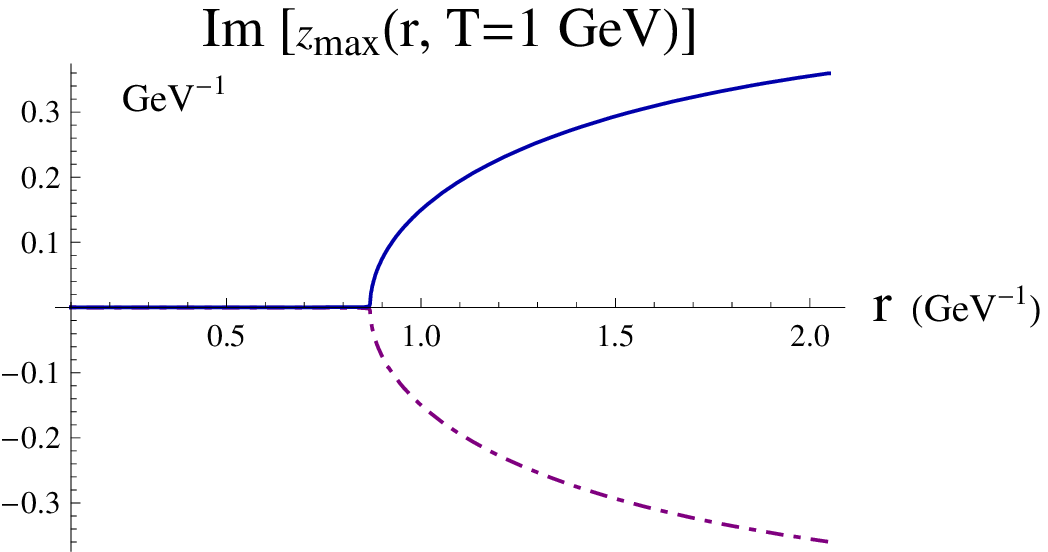}
\caption{Real and imaginary parts of the roots of \eq{zmax} plotted as a function of
  quark--anti-quark separation $r$.~\vspace*{-2mm}}
\label{fig1}
\end{figure}

First we note that the extension of the solution for $z_{max}$
following from \eq{zmax} to $r > r_c$ is not unique. Two most
important roots of \eq{zmax} found in
\cite{Rey:1998bq,Brandhuber:1998bs} are shown in \fig{fig1}, which
depicts real and imaginary parts of $z_{max}$ as functions of
separation $r$. There are other roots of \eq{zmax} that are not shown
in \fig{fig1}: they are either negatives of the roots in \fig{fig1} or
the complex conjugates of the roots in \fig{fig1} and of their
negatives. Such extra roots lead to physically irrelevant
configurations and are not shown in \fig{fig1}. To determine
  which root gives the correct potential from first principles one has
  to (at least) calculate quantum $o(1/\sqrt{\lambda})$ corrections to
  the quasi-classical results shown above. While such a calculation is
  necessary, it would be rather tedious and is left for future work.
  Here we will demand that the correct root maps on the Maldacena
  vacuum solution \cite{Maldacena:1998im} in the zero temperature
  limit. In addition we will impose unitarity to single out the right
  root.

Of the two roots shown in \fig{fig1} only one (denoted by the solid
line) maps onto Maldacena's solution behavior of $z_{max} = r c_0$ at
small $r$ \cite{Maldacena:1998im}. Since, on physical grounds, we want
our potential to recover the zero temperature result
\cite{Maldacena:1998im} at small $r$ we will keep this root, and
discard the other root denoted by the dash-dotted line in \fig{fig1}.
The root given by the solid line develops a positive imaginary part
for $r > r_c$, as shown in the bottom portion of \fig{fig1}. As the
complex conjugate of this root would also be a solution of \eq{zmax},
while mapping onto Maldacena's solution for small $r$, the question
arises about the choice of one root over its complex conjugate. To
select the root we note that the quantum-mechanical time-evolution
operator in Minkowski metric is $e^{- i E \, t} \sim e^{\text{Im} [E]
  \, t}$.  Demanding that the probability of a state does not exceed
one we obtain $\text{Im} [E] < 0$, leading to $\text{Im} [V(r)] < 0$.
This condition allows us to single out the solid line in \fig{fig1}
over its complex conjugate as the physically relevant root.
\begin{figure}[h]
  \centering 
\includegraphics[width=7.cm]{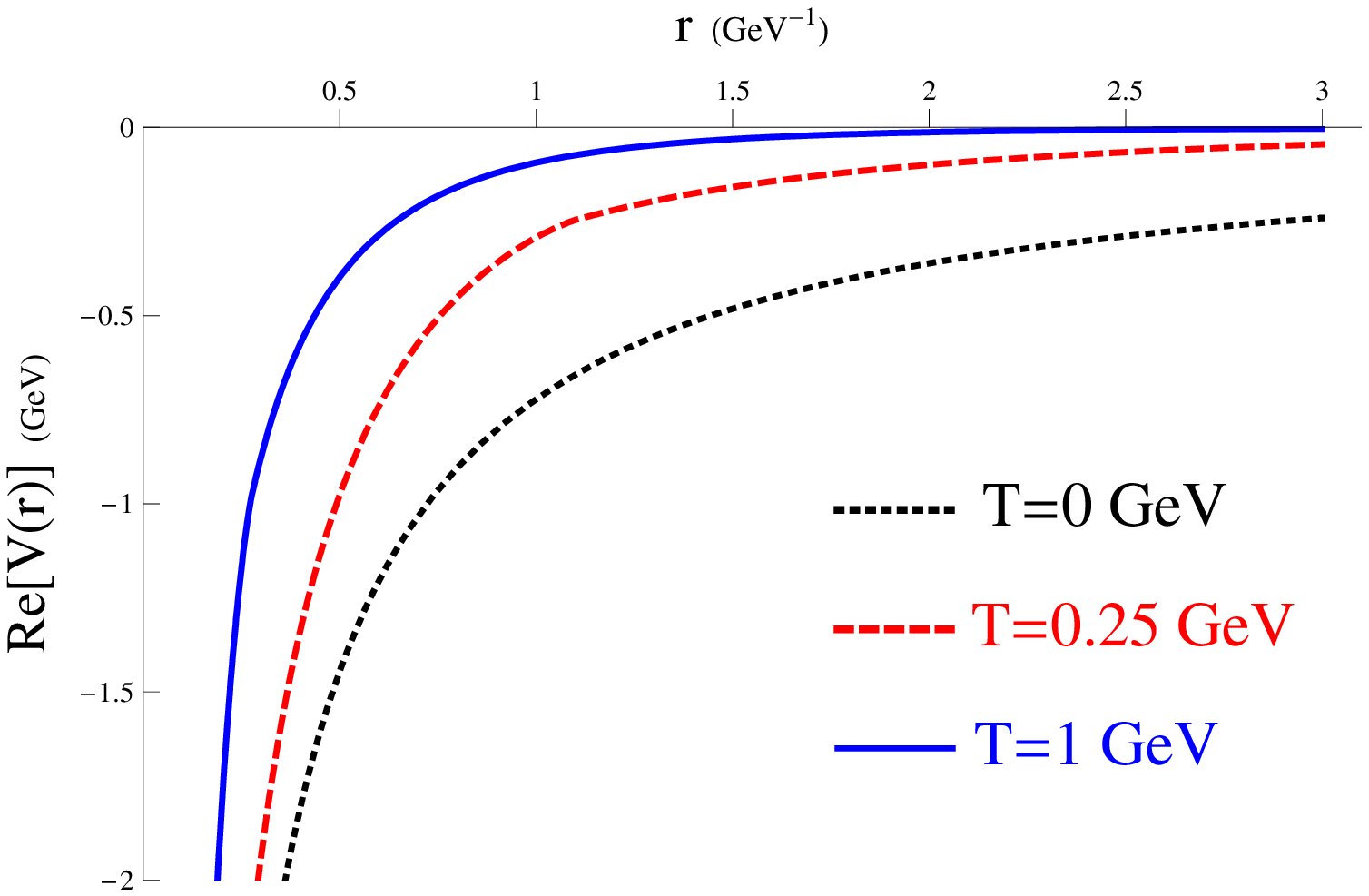}
\includegraphics[width=7.cm]{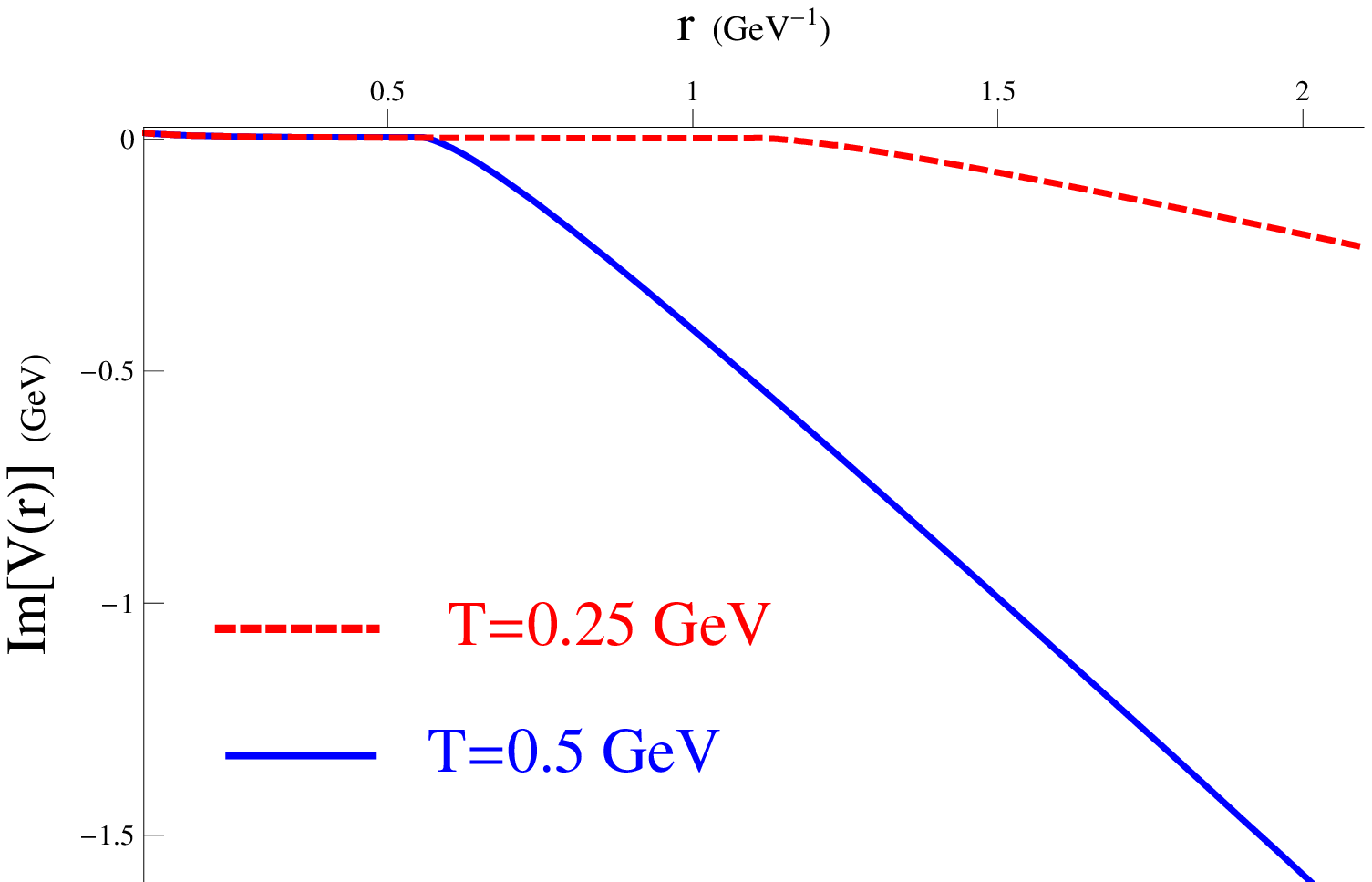}
\caption{The real and imaginary parts of the heavy quark potential 
  plotted as functions of the separation $r$ for several different
  temperatures. We put $\lambda =10$.~\vspace*{-3mm}}
\label{fig2}
\end{figure}

Using the solid line root of \fig{fig1} in \eq{pot}, we can plot the
real and imaginary parts of the resulting potential. The plots are
shown in \fig{fig2}. In the top panel of \fig{fig2} we show the real
part of the heavy quark potential for two non-zero temperatures, along
with the zero-T curve for comparison. One can see that the non-zero
temperature curves are indeed strongly screened compared to the zero
temperature case, but remain non-zero at all $r$. The subtraction
scheme proposed above in \eq{subtr} insures that $\text{Re} [V(r)]$
given by \eq{pot} approaches zero at large $r$. The use of the
subtraction scheme proposed in \cite{Rey:1998bq,Brandhuber:1998bs}
would have led to $\text{Re} [V(r)]$ going to a positive constant as
$r \rightarrow \infty$.

As is clear from the lower panel in \fig{fig2}, the heavy quark
potential develops an imaginary part for $r > r_c$. This means the
potential becomes absorptive, as the $q\bar q$ singlet state may melt
in the medium. The rate of absorption increases with $r$, as larger
pairs are more likely to decay.  The existence of an imaginary part in
the heavy quark potential has been previously observed in perturbation
theory in \cite{Laine:2006ns,Brambilla:2008cx}.  While $\text{Re}
[V(r)]$ in \fig{fig2} does not have a kink, there is a region near
$r=r_c$ where the slope of the curve changes very fast.  This rapid
change is due to the potential developing an imaginary part, which
should quickly reduce the force on the quarks.

As one can explicitly check from \eq{pot}, at small $r$ we recover the
zero temperature potential of \cite{Maldacena:1998im}
\begin{align}\label{smallr}
  V (r) \bigg|_{r \, T \ll 1} \approx - \frac{\sqrt{\lambda}}{2 \, \pi
    \, c_0^2 \, r}. 
\end{align}

At large separations $r$ we first use \eq{zmax} to write
\begin{align}\label{zexp}
  z_{max} \bigg|_{r \, T \gg 1} = \frac{r}{\pi \, c_0} +
  \frac{1+i}{\pi \, c_0^2} \, z_h + \frac{\pi^3 \, c_0^3}{3} \,
  \frac{z_h^4}{r^3} \notag \\ - (1+i) \, \frac{6}{5} \, \pi^3 \, c_0^2
  \, \frac{z_h^5}{r^4} + o \left( \frac{z_h^6}{r^5} \right).
\end{align}
Using \eq{zexp} in \eq{pot} we obtain
\begin{align}\label{rvexp}
  \text{Re} [V (r)]\bigg|_{r \, T \gg 1} \, = \, - \frac{\pi^3 \,
    c_0^3}{4} \, \sqrt{\lambda} \, \frac{z_h^3}{r^4} + o \left(
    \frac{z_h^4}{r^5} \right)
\end{align}
and
\begin{align}\label{zvexp}
  \text{Im} [V (r)] \bigg|_{r \, T \gg 1} \! \! =  -
  \frac{\sqrt{\lambda}}{\pi} \, \frac{1}{2 \, z_h} \, \left[
    \frac{r}{z_h} - \frac{1}{c_0} + o \left( \frac{z_h}{r} \right)
  \right]. 
\end{align}
As one can see from \eq{rvexp}, instead of the exponential falloff
with $r$ characteristic of Debye screening, which would have been
expected from small coupling perturbation theory and which was
postulated for ${\cal N} =4$ SYM theory at strong coupling in
\cite{Bak:2007fk}, the real part of the heavy quark potential falls
off as a power, $\text{Re} [V (r)] \sim 1/T^3 r^4$, at large $r$. If
our hypothesis of using the complex string configurations is
confirmed, this would be an interesting new type of screening for the
potential. However, the large negative imaginary part of the potential
(\ref{zvexp}) leads to exponential decay with time of the initial
(built in by construction) color correlation between the quark and the
anti-quark in the Wilson loop.

Combining the large- and small-$r$ asymptotics in Eqs. (\ref{smallr}),
(\ref{rvexp}) we can interpolate the real part of the potential to
write an approximate formula
\begin{align}\label{approx}
  \text{Re} [V (r)] \approx - \frac{\sqrt{\lambda}}{2 \, \pi \, c_0^2
    \, r} \, \frac{r_0^3}{(r_0 + r)^3}
\end{align}
with the new scale $r_0$ equal to
\begin{align}\label{r0}
  r_0 \, = \, z_h \, \pi \, c_0 \, \left( \frac{\pi \, c_0^2}{2}
  \right)^{1/3} \, \approx \, \frac{2.702}{\pi \, T}.
\end{align}
\eq{approx} fits the curves on the upper panel of \fig{fig2} quite
well. The parameter $r_0$, defined by \eq{approx} and given in
\eq{r0}, can be interpreted as the screening length.

While our power-law screening $\text{Re} [V (r)] \sim 1/T^3 r^4$ is
very different from the exponential falloff due to Debye screening
$\text{Re} [V (r)] \sim \exp (- m_D \, r)$ in coordinate space, the
difference is not so profound in momentum space. Define a Fourier
transform of the potential
\begin{align}
  {\tilde V} (q) \, = \, \int d^3 r \, e^{- i {\vec q} \cdot {\vec r}}
  \, V(r). 
\end{align}
Using the small-$r$ asymptotics (\ref{smallr}) 
one can easily show that at large $q = |{\vec q}|$ the
Fourier transform of the potential scales as ${\tilde V} (q) \sim
1/q^2$, in agreement with the standard perturbative result. At small
$q$
\begin{align}
  {\tilde V} (q=0) \, = \, \int d^3 r \, V(r) \, \sim \, r_0^2 \, \sim
  \, \frac{1}{T^2}.
\end{align}
Such asymptotic behavior is very similar to the Debye screening in the
infrared (IR), given by the screened propagator ${\tilde V}_D (q) \sim
1/(q^2 + m_D^2)$ in the perturbation theory with Debye mass $m_D \sim
T \sim 1/r_0$. Hence our ${\tilde V} (q)$ has qualitatively the same
UV and IR asymptotics as the standard perturbative Debye-screened
potential. The main difference is in the shape of ${\tilde V} (q)$ at
finite $q$: unlike ${\tilde V}_D (q)$ our ${\tilde V} (q)$ is concave
at all $q$.

To summarize, we proposed a method of calculating the heavy quark
potential in the finite-T strongly-coupled ${\cal N} =4$ SYM theory in
the region of separations $r$ and/or temperatures $T$ where the
classical string configuration does not exist. We used an analogue of
the complex trajectories method in the quasi-classical quantum
mechanics \cite{LL3} and analytically continued the string
configurations into the region of complex coordinates. This allowed us
to obtain a potential which is physically meaningful for all values of
$r$ and $T$. The potential develops an imaginary absorptive part. We
would like to stress that instead of Debye screening at large
separations the real part of the potential falls off as a power of
the separation, which is a new and never before observed phenomenon in
relativistic quantum field theories.

~\hspace*{1mm}

We would like to thank Eric Braaten, Dick Furnstahl, Andreas Karch,
Samir Mathur and Larry Yaffe for informative discussions.

This research is sponsored in part by the U.S. Department of Energy
under Grant No. DE-FG02-05ER41377.





\end{document}